An Explanation of the Missing Flux from Boyajian's Mysterious Star.

Peter Foukal, Nahant, MA, 01908, USA

Abstract

A previously unremarked star in the constellation Cygnus has, in the past year, become known as the most mysterious object in our Galaxy. Boyajian's star exhibits puzzling episodes of sporadic, deep dimming discovered in photometry with the Kepler Mission. Proposed explanations have focused on its obscuration by colliding exo- planets, exo- comets, and even intervention of alien intelligence. These hypotheses have considered only phenomena external to the star because the radiative flux missing in the dimmings was believed to exceed the star's storage capacity. We point out that modeling of variations in solar luminosity indicates that convective stars can store the required fluxes. It also suggests explanations for a) a reported time – profile asymmetry of the short, deep dimmings and b) a slower, decadal scale dimming reported from archival and Kepler photometry. Our findings suggest a broader range of explanations of Boyajian's star, that may produce new insights into stellar magneto-convection.

I. Introduction

Boyajian's star is a main sequence F 2 object whose peculiar light curve was noted in photometry obtained with the Kepler Mission, by the Planet Hunter volunteers. The most interesting aspect of the recorded four- year light curve are episodes of sporadic dimming by as much as 20 %, each lasting approximately a few to tens of days, separated by periods of relatively quiescent output lasting typically 100's of days (Boyajian et al. 2016; Schaefer 2017)

These sporadic dimmings have attracted wide attention and Boyajian's star has been described as the most mysterious star in our Galaxy. Proposed explanations include obscuration of the star by colliding exo- planets, exo- comets, and even proposals involving alien intelligence (see Wright & Sigurdson 2016 for an overview). These explanations external to the star proceed from the assumption that the missing radiative flux during a dimming cannot be stored within the star itself. This problem of accounting for the missing flux has assumed sufficient importance that the object is also known as the "where's the flux" (wtf) star (Boyajian et al. 2016).

The purpose of this Letter is to point out that, provided the star possesses a convection zone immediately below the photosphere, its missing flux can be stored as a small increase of its internal and potential energy. This surprising ability of a convective star to store heat flux blocked from emergence as photospheric radiation has been demonstrated in the case of the sun.

II. Storage of Heat Flux Obstructed by Sunspots

In the solar case, variations in the luminosity are produced by changes in the photospheric area covered by magnetic dark spots (e.g. Willson et al. 1981; Foukal et al. 2006). The radiative flux missing from them poses a similar problem to that encountered with Boyajian's star. The fluxes are smaller and the type of obstruction may differ, but as shown below, the explanation is likely to be similar.

The storage of perturbations to solar photospheric heat flow was studied, after discovery of variation in total solar irradiance in 1981, using time dependent models of heat flow around thermal plugs representing spots (Spruit 1982a,b; Foukal, Fowler & Livshits 1983). The results were instructive. One might think that heat flow would re-adjust quickly around such a thermal plug, with little influence on the luminosity. But instead the diverted heat is rapidly (in days) redistributed throughout the solar convection zone by the high thermal diffusivity of solar convection. It is stored as a very small global heating and expansion of that envelope, with negligible influence on the photospheric temperature and luminosity. Since this blocked heat does not immediately re-emerge anywhere to compensate for the local blocking, the solar luminosity simply decreases in proportion to the area of the obstruction, as is observed in radiometry of the solar irradiance.

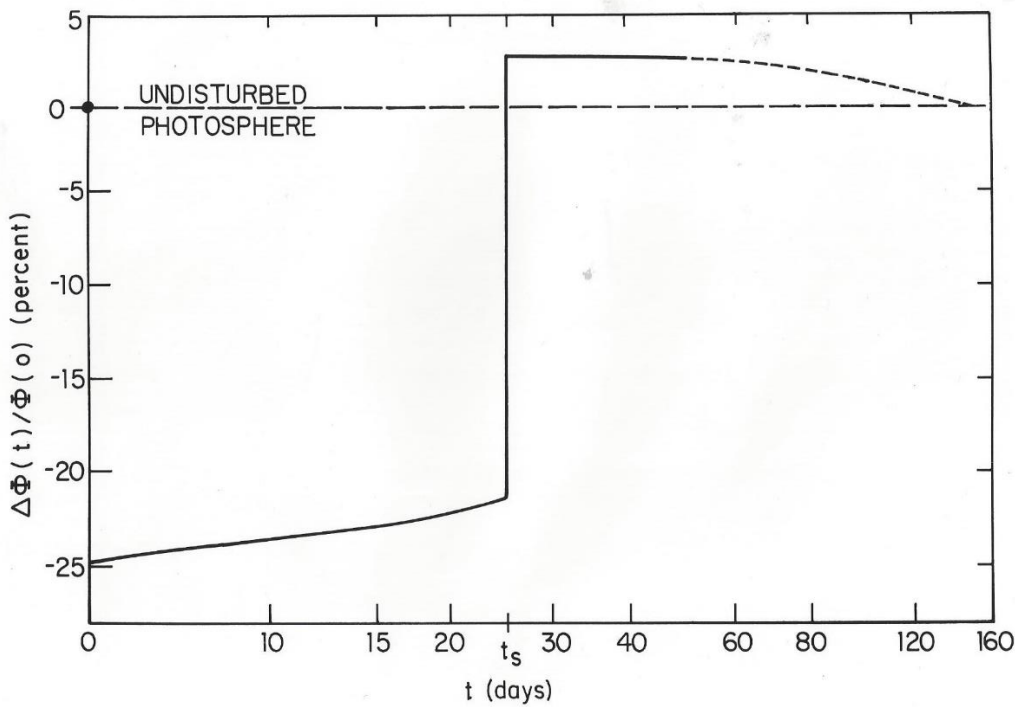

Fig.1 Time dependence of radiative flux $\Phi(t)$ after insertion of an obstruction of $10^3$ km diameter at t = 0, and its removal at t = ts, in a convection zone of depth $10^4$ km (from Foukal, Fowler, & Livshits 1983).

When the plug is removed, the solar luminosity slowly relaxes back to its original value. The surprise in these calculations was the very long time required; the radiative relaxation time of the convection zone, or about 100,000 yrs. So the storage is essentially perfect. The calculations were carried out using a simple mixing length model of solar convection. But the result depends little on the details of the convection. The main point is that the radiative relaxation time of the solar convection zone is about $10^7$ times longer than its internal heat diffusion time.

III. Application of this Model to Boyajian's Star

The behavior of the radiative flux is illustrated in Figure 1, for a heat flow obstruction inserted into a shallower convection zone than the sun's. Immediately after the insertion at t=0, the luminosity drops by an amount equal to the product of the obstruction's area and photometric contrast, but begins to rise right away because the shallow convection zone does not store the blocked heat as perfectly as does the sun's. When the obstruction is removed at $t = t_s$ the slightly heated and expanded photosphere radiates at a higher luminosity, which decays relatively quickly in this case, back to the original value at t = 0.

The gradual dimming seen at $t > t_s$ (N.B. the time scale in Figure 1 is logarithmic) might explain the slowly decreasing brightness of Boyajian's star reported from studies of the Harvard archival plate collection and of the Kepler data (Schaefer 2017; Montet & Simon 2016). These studies provide evidence that, besides the occasional deep dips, the star's luminosity decreased gradually by about 15% over the last century, and by 3 % over the 4- year extent of the Kepler observations.

Either of these rates would be consistent with a shallower convection zone than that of the sun, since the zone's radiative relaxation time scales roughly as its depth. The reported rates of between 0.1 – 1 %/yr would correspond to convection zone depths of roughly 1-5 x $10^4$ km (Foukal, Fowler & Livshits 1983), although these estimates are very approximate because they depend on the depth structure of the density and specific heat, which is different in the sun and an F star. This suggests that such a star would be constantly dimming from a state of enhanced luminosity caused by the sporadic flux blocking episodes.

The flux increase that sets in immediately after t = 0 in Figure 1 should distort the observed time profile of luminosity during the deep dips by increasing the rate of flux recovery. The time profile shown in Figure 1 shows the response of luminosity to an obscuration whose time profile is a box function initiated at t = 0 and removed at $t = t_s$. The observed profile will be its convolution with the time profile of an obscuration developing and decaying on the star. If this were a delta function in time, then the observed profile would be that of Figure 1. More generally, it will be asymmetric, with a more rapid egress than ingress because of the upward slope seen between t = 0 and $t = t_s$ in Figure 1. It is interesting that the most nearly symmetric of the observed dips (Dip 5; Boyajian et al. 2016), exhibits such a faster egress than ingress.

The illustration presented here is based on the case of obstruction by a spot because study of sunspot influence on solar irradiance has already provided solid information on this mechanism. But as we discuss below, other kinds of obstructions to radiative flux may be more likely in an F star. The basic features of the luminosity's time behavior should, however, be similar to those outlined in Figure 1.

The solar luminosity variation is at the 0.1% level, rather than the 20% observed in Boyajian's star. But the same explanation has been applied to other late type variable stars, like the RS CVn and BY Dra stars, whose huge star - spots can cause flux dips of similar size to those observed in Boyajian's star. That explanation, which employed a polytropic approximation to the star's envelope, showed that the effect of even such huge spots should be small (Spruit 1982a,b). For instance, the radius increase associated with the storage is only a small fraction of the photospheric scale height.

IV. What Causes the Heat Flux Obstruction on Boyajian's Star?

The result presented above depends little on the nature of the thermal perturbation. Our simulations represented a spot simply as a shallow cylinder of reduced eddy thermal conductivity. But change in the F2 star's near - photospheric chemical composition, for instance, might change its opacity structure and $T_{eff}$, producing a similar effect.

So the dimmings of Boyajian's star do not require magnetic activity, although there is some evidence of a high frequency photometric variation of about 500 ppm amplitude with a period of about 0.88 days, which might be caused by spots and faculae rotating at this rate (Boyajian et al 2016). According to these authors, this short period agrees with the rotational velocity v sin i derived spectroscopically (which is several times higher e.g. than those tabulated for a sample of 22 rapidly rotating Kepler F stars studied by Mathur et al. 2014). An alternative interpretation (Makarov & Goldin 2016) of this high frequency variation in terms of noise in the Kepler data (that happens to agree with the spectroscopic rate) seems less likely but deserves further investigation.

The most important proviso is that this F2 star should possess a substantial convection zone immediately below the photosphere. This is required to achieve the high thermal diffusivity that makes the sun's convection zone essentially a thermal "superconductor" (Spruit 1982a). Convection is considered to persist in late-type stars up to Teff ~ 7500 K (Bohm Vitense 1992) well above the 6750K temperature of Boyajian's star . But demonstration that the 0.88 d variation is caused by magnetic activity would help to support its existence.

If the star is convective and magnetically active, the recent modeling of F star dynamos (Augustson et al. 2013) might be of interest. The chaotic behavior of the sun's dynamo causes a centennial- scale amplitude modulation of its 11 yr magnetic cycle, best known for its most recent "Maunder" minimum, when spots essentially disappeared between about 1645 and 1715. The F star modeling suggests that the 30 times higher rotation rate of this star might produce the sporadic variations in the field (and its associated luminosity modulation) on the shorter time scale that is observed in the Kepler photometry. The activity seen in the models lies mainly in a low latitude belt, relatively uniformly distributed in longitude. So its rapid 0.88 d. rotation might produce only a small photometric modulation, consistent with the 500 ppm variation observed in the Kepler data.

More study is required to understand why, of all the Kepler Mission stars analyzed, only this one exhibits such odd behavior. Observational discrimination against F2 stars may be a factor, since F2 lies near the "hot" limit of the Kepler Mission stars, which were chosen for their likelihood of harboring life-supporting planets.

The possibility that the star lies near the cut-off for convection might also play a role. Certainly, the modeling shows that both an F star's magnetic field and the shear associated with its differential rotation can make it difficult for convection to transport the star's heat flux. This would increase the likelihood of random variations in convective efficiency. Such variations might arise from changes in the number and geometry of convective cells (e.g. Hartmann & Rosner !979). So Boyajian's star might present us with a fascinating example of a star "sputtering" to get its heat flux out.

V. Summary and Conclusions

Our understanding of solar luminosity variation indicates that Boyajian's star might store its missing radiative fluxes internally like other late – type stars whose photospheric flux is blocked by obstructions of magnetic or other nature. The sporadic dimming, the slower flux decrease over an annual time scale, and even the asymmetry of the dips, appear to be explainable in this way. There seems to be little need to postulate external explanations of the missing flux, on these grounds.

Further work is required to identify the cause of the radiative flux obstruction. Magnetic activity, differential rotation, sporadic changes in photospheric abundances, and simply random variation in convective efficiency are all candidates worth investigating.

A full understanding of Boyajian's star requires an answer to why only one example has been found. The low likelihood of the scenarios postulated in the external explanations suggested so far is specifically aimed at addressing this question. But these complex models encounter serious difficulties in explaining the observed phenomena (e.g. Schaefer 2017). Their merits must be weighed against the acceptance of internal storage as the basis of our understanding the photometric variability of late type stars at amplitudes ranging from that of the sun to that of RS Cvn objects. It seems logical to look to a variant of that established property, as explored here, to explain the behavior of Boyajian's star as well.

I thank Tabetha Boyajian, Brad Schaefer and Mark Giampapa for comments on a draft of this paper. I am grateful to Mark Miesch for drawing my attention to the MHD modeling results and the impact of strong differential rotation on stellar heat transport efficiency.